\documentclass[%
 aip,%
 apl,
 amsmath,amssymb,%
 reprint,%
 floatfix,%
]{revtex4-2}%
\usepackage{graphicx}
\usepackage{dcolumn}
\usepackage{bm}
\usepackage[utf8]{inputenc}%
\usepackage[T1]{fontenc}%
\usepackage{mathptmx}%
\usepackage{etoolbox}%
\usepackage[skip=-1pt, singlelinecheck=off]{subcaption}%
\usepackage{nicefrac}
\begin{document}%
\preprint{AIP/123-QED}%
\title[Title]{Hybrid ultrathin metasurface for broadband sound absorption}%
\author{Marnix P. Abrahams }%
\affiliation{%
Universit\'e de Lorraine, CNRS, Institut Jean Lamour, F-54000 Nancy, France
}%
\author{Mourad Oudich}%
 \email[Authors to whom correspondence should be addressed: ]{mourad.oudich@univ-lorraine.fr and badreddine.assouar@univ-lorraine.fr}%
\affiliation{%
Universit\'e de Lorraine, CNRS, Institut Jean Lamour, F-54000 Nancy, France
}%
\affiliation{
Graduate Program in Acoustics, The Pennsylvania State University, University Park, Pennsylvania 16802, USA
}%
\author{Yann Revalor }%
\affiliation{%
Dassault Aviation, 92552 St-Cloud, France%
}%
\author{Nicolas Vukadinovic }%
\affiliation{%
Dassault Aviation, 92552 St-Cloud, France%
}%
\author{Badreddine Assouar}%
 \email[Authors to whom correspondence should be addressed: ]{mourad.oudich@univ-lorraine.fr and badreddine.assouar@univ-lorraine.fr}%
\affiliation{%
Universit\'e de Lorraine, CNRS, Institut Jean Lamour, F-54000 Nancy, France
}%

\date{\today}
%
\begin{abstract}%
To this day, achieving broadband low-frequency sound absorption remains a challenge even with the possibilities promised by the advent of metamaterials and metasurfaces, especially when size and structural restrictions exist. Solving this engineering challenge relies on stringent impedance matching and coupling of the multiple independent local resonators in metasurface absorbers. In this letter, we present an innovative design approach to broaden the sound absorption bandwidth at low-frequency regime. A hybrid metasurface design is proposed where four coupled planar coiled resonators are also coupled to a %
well designed thin planar cavity. This hybrid metasurface creates a broad sound absorption band (130-200 Hz) that is twice as wide as that of the traditional single layer metasurface utilizing four coiled cavities at a deep sub-wavelength thickness ($<\ \nicefrac{\lambda}{51}$). This design strategy open routes towards engineering a class of high performance thin metasurfaces for ultra-broadband sound absorption while keeping the planar size unchanged.

\end{abstract}%
\maketitle%
%
%
\begin{figure*}[t!]%
\begin{subfigure}{0\textwidth}%
\phantomcaption\label{fig:fig1a}%
\end{subfigure}%
\begin{subfigure}{0\textwidth}%
\phantomcaption\label{fig:fig1b}%
\end{subfigure}%
\begin{subfigure}{0\textwidth}%
\phantomcaption\label{fig:fig1c}%
\end{subfigure}%
    \includegraphics[scale=1]{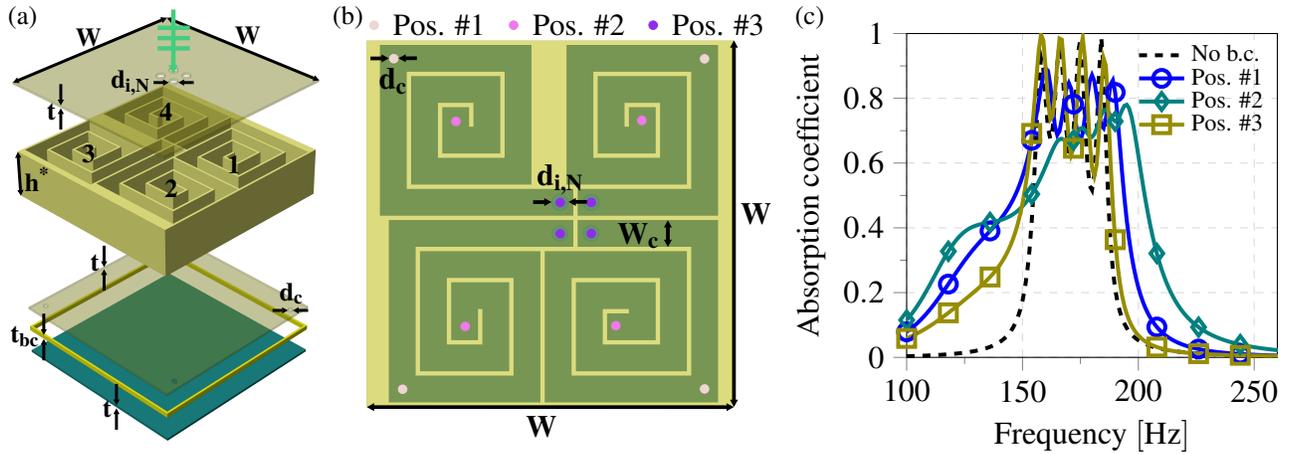}%
\caption{\label{fig:fig1} %
(a) Hybrid meta-surface made of four coiled resonators and a planar back-cavity. A perforated plate with four holes is located between the coiled resonators and the back cavity where each hole connects each to the back cavity. The coiled (unit-cells) are numbered 1 through 4 and consist of an inlet hole and a coil, its backplate is perforated to allow flow through to the shared back-cavity.
(b) Top-view of the metasurface indicating three different configurations for the positions of the holes connecting the coiled resonators to the back-cavity. These positions are labeled Pos.\#1, Pos.\#2, and Pos.\#3. The values of the geometrical parameters in (a) and  (b) are provided in Table \ref{tab:table1}. (c) The sound absorption coefficient for the three connecting holes configurations presented in solid line with symbol : Pos.\#1 (blue circle), Pos.\#2 (green diamond), and Pos.\#3 (olive square), with the hole diameter $d_c$ being 0.85 mm. The dashed line corresponds to the four coiled metasurface without back cavity. (Symbols for illustration purposes only.)
}%
\end{figure*}%
%
%
%
\begin{table*}[t!]%
\caption{\label{tab:table1}The design parameters depicted in Fig. \ref{fig:fig1a} \& \ref{fig:fig1b} \& Fig. \ref{fig:fig3a} %
being static unless locally otherwise specified.}%
\begin{ruledtabular}%
\begin{tabular}{ccccccccccc}%
$\text{d}_{\text{i,N}}$ [mm]& $\text{L}_{\text{1}}$ [mm]& $\text{L}_{\text{2}}$ [mm]& $\text{L}_{\text{3}}$ [mm]& $\text{L}_{\text{4}}$ [mm]& $\text{W}$ [mm]& $\text{h}^{*}$ [mm]& $\text{W}_{\text{c}}$ [mm]& $\text{d}_{\text{c}}$ [mm]& $\text{t}$ [mm]& $\text{t}_{\text{bc}}$ [mm]\\%
\hline
7 & 460 & 430 & 410 & 380 & 165 & 40 & 12 & 1 & 2 & 4
\end{tabular}%
\end{ruledtabular}%
\end{table*}%
%
%
    The creation of thin rationally designed materials, namely metasurfaces\cite{Assouar2018}, capable of high sound absorption with broadband capability at a deep subwavelength scale is highly desirable for a wide range of applications : room acoustics, automobiles, aerospace etc.  In traditional sound-absorbing materials such as foam, fiber glass and mineral wool, the relationship between the wavelength and the material thickness constrains the absorption spectrum. This makes decreasing the material thickness difficult while maintaining the desired sound absorption performance at low frequency\cite{Yang2017_2}. Fortunately, the advent of acoustic metamaterials has reshaped the entire field of acoustics and particularly the sound absorption with the emergence of ultrathin metasurfaces capable of deep-subwavelength high sound absorption\cite{Cai2014, Jimnez2016, Li2016, Peng2018, Donda2019, Zhu2019, Ji2020, Ma2023, Wu2022}.  %
       		
	Still, the research within the field of acoustic metasurfaces shows that the realization of a broadband high absorption rate at low-frequency regime remains challenging, with the thickness being the leading limiting factor\cite{Yang2017, Zhou2021}. %
	One of the initial strategies adopted for the design of sound absorbing metasurfaces is the utilization of thin membranes distributed in a rigid grid panel\cite{Mei2012, Yang2015, Ma2014, Guo2022} . However, this type of thin resonating structure relies on mastering the mechanical tension of the membranes which poses challenges in fabrication. A different design methodology consists on combining a perforated panel with a resonator coiled in a planar configuration such that the total thickness of the whole structure can be reduced in the third dimension ($z$-direction)\cite{Liang2012, Cai2014, Li2016, Huang2018, Ma2023} . This design strategy enables perfect absorption at deep subwavelength scale, but with a narrow frequency band corresponding to the resonance frequency of the structure. Optimization methods on the resonant cavity design were conducted to drive the high sound absorption capability into low frequencies using internal patterning\cite{Donda2019, Donda2022}, or embedded and shaped neck\cite{Huang2019, Guo2020, Bi2023, Zhang2022, Duan2020}. 

    Broadband sound absorption capability was later demonstrated with a super-cell encompassing multiple unit-cells with coiled cavities. The resonance frequencies of the latter were chosen judiciously to enable their mutual coupling and achieve broadband high absorption rate  \cite{Ji2020, Peng2018, Donda2019, Guo2022, Wu2022, Huang2020}. Recently, nonlocal coupling has been considered as a route for broadening the absorption band \cite{Zhou2021, Zhu2021, Wang2023}. In addition to coupling nearest-neighbor resonators, nonlocality consists of coupling the cavities of second and higher order neighboring cavities to broaden further the bandwidth. These strategies for broadening the bandwidth of high absorption are mostly based on coupling resonant units where a supercell has to be constructed with multiple units which increases the size of the absorber in the planar direction. In other words, the bandwidth interval strongly depends on the number of unit cells which limits the metasurface practicability. In this letter, we aim on overcoming this limitation for coiling-up space geometry cavity-based metasurfaces. 
 
 Starting from a system of four coupled coiled cavities \cite{Li2016, Donda2019}, we present an innovative design solution to broaden the sound absorption bandwidth beyond what is allowed by the classical coupled cavities and without change in the planar size of the metasurface. To do so, a hybrid concepts is introduced where the four coupled coiled resonators are also coupled to a rationally designed thin planar back cavity in the normal direction (Fig. \ref{fig:fig1a}). We will demonstrate that judicious design of the planar back cavity enables an enlargement of the bandwidth by a factor of two. We aim for a design that one can integrate in the airplane fuselage wall to significantly mitigate the motor-noise normally heard by the passengers, which ranges from 100 to 500 Hz. Our proposed absorbing metasurface design uses non-local coupling for wide-band absorption below 200 Hz with a sub-wavelength thickness of 50 mm ($\nicefrac{\lambda}{51}$). 
 
%
%
    We attain our design using the formulas (2 \& 5) from Li and Assouar\cite{Li2016} from which we create an optimized super-cell %
    with equally spaced unitary absorption peaks between 150 and 180 Hz. %
    Making use of the commercial numerical simulation software Comsol Multiphysics, we further optimize the design of the metasurface. Figure 1(a) illustrates our hybrid metasurface that combines the four coiled resonators with a perforated plate and a back-cavity that were placed between the coils and back rigid wall. This added back cavity is shared between the four unit-cells and connected with four holes of the same diameter ($d_c$). The geometrical parameters of the hybrid metasurface are indicated in Fig. \ref{fig:fig1a}, and their values are provided in Table \ref{tab:table1} which are adopted throughout the entire letter unless otherwise specified.%

%
\begin{figure*}[t!]
\begin{subfigure}{0\textwidth}%
\phantomcaption\label{fig:fig2a}%
\end{subfigure}%
\begin{subfigure}{0\textwidth}%
\phantomcaption\label{fig:fig2b}%
\end{subfigure}%
    \includegraphics[scale=1]{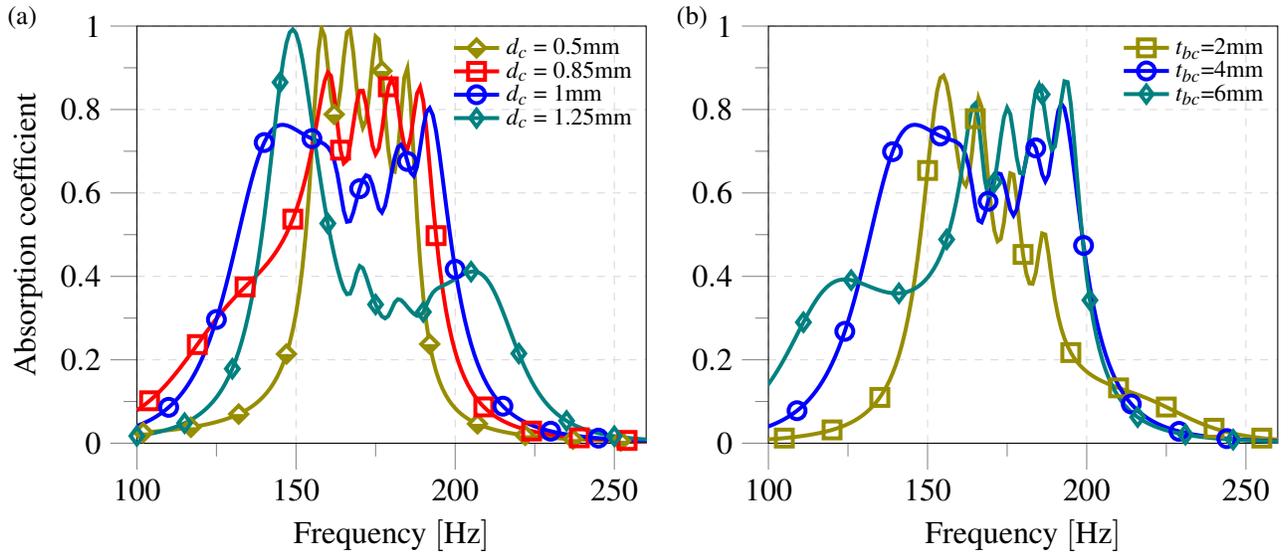}%
\caption{\label{fig:fig2} %
(a) The sound absorption coefficient dependence on the hole diameter (\text{$d_c$}) of the connecting holes in the design displayed in Fig. \ref{fig:fig1a} for the holes location ''Pos. $\#1$''. (b) The absorption dependence on back-cavity thickness (\text{$t_{bc}$}) when the connecting hole diameter $d_c$ is equal to 1 mm and the total thickness is kept at 50 mm while altering the thickness $\text{h}^\text{*}$ through the formula $\text{h}^\text{*}=50\text{[mm]}-3\text{t}-\text{t}_\text{bc}$.  (Symbols for illustration purposes only.)
}%
\end{figure*}%

In our design optimization, it is important to realize that the holes connecting the coiled cavities to the back-cavity (connecting holes) can be located at any position along the length of each coiling, which divides the channels of the coiled resonator into different resonating domains thereby greatly affecting the absorption behavior. In order to find the optimal location, an optimization process was conducted, and three of the possibilities for the connecting holes localization considered are shown: at the corners (Position 1), at the end of the coiled cavity (Position 2), and at the same position as the inlet holes (first end of the cavity) (Position 3). These positions are illustrated in Fig. \ref{fig:fig1b} and labeled Pos.\#1, Pos.\#2 and Pos.\#3. 

Figure \ref{fig:fig1c} presents the sound absorption rates for different configurations of the metasurface. The dashed line corresponds to the classical metasurface with the four coiled cavities without a back cavity nor its adjoining perforated plate. The three solid lines with markers correspond to the hybrid metasurface which includes the back cavity with its perforated plate. These three curves are associated with the three considered positions of the connecting holes as described previously: circular marker for Pos.\#1, diamond for Pos.\#2, and square for Pos.\#3. The case of the classical metasurface (dashed line) (w/o back cavity) exhibits near unitary absorption rate at four frequencies equally spaced between 150-180 Hz and the coupled coiled resonator super-cell acquire a 32 Hz bandwidth at 50\% absorption (155-187 Hz). %
When adding the back cavity with the perforated panel, two observations can be made. First,  the absorption bandwidth increases no matter the location of the connecting holes (solid line curves in comparison with the dashed line curve) by between 5 to 16 Hz. Secondly, the absorption coefficient is reduced when considering positions \#2 and \#3 of the connecting holes while the bandwidth increases. The enlargement of the bandwidth of high absorption is made at the cost of not reaching unitary absorption rate. Hence, there is a trade-off between the absorption bandwidth and coefficient: the increase of the absorption band causes a reduction of the absorption coefficient. However, even if perfect absorption could not be attained when widening the bandwidth, the hybrid metasurface still provides high absorption coefficient of at least 70\% (Fig. 1c). Taking this trade-off into account alongside our objective, we argue that the location of the connecting holes at `Pos. \#1' is comparatively the best scenario.

%
Considering `Pos. \#1' for the connecting holes, we have also conducted further optimization on their diameter and the back-cavity thickness, and the results are illustrated in Fig. \ref{fig:fig2}. %
In figure \ref{fig:fig2a}, the increase of the diameter increases the bandwidth significantly while the absorption coefficient is reduced. Also from \ref{fig:fig2b}, it is clear that the thickness of the back cavity has a significant impact on the absorption bandwidth and rate. 
Considering the trade-off evident throughout these parametric optimizations, we state the optimal 
diameter for the connecting holes and the back cavity thickness as $d_c=1mm$ and $t_{bc}=4mm$ respectively, leading to an absorption coefficient above 50\% for a bandwidth of 70 Hz (130-200 Hz). %

%
%
\begin{figure*}[t!]
\begin{subfigure}{0\textwidth}%
\phantomcaption\label{fig:fig3a}%
\end{subfigure}%
\begin{subfigure}{0\textwidth}%
\phantomcaption\label{fig:fig3b}%
\end{subfigure}%
\begin{subfigure}{0\textwidth}%
\phantomcaption\label{fig:fig3c}%
\end{subfigure}%
    \includegraphics[scale=1]{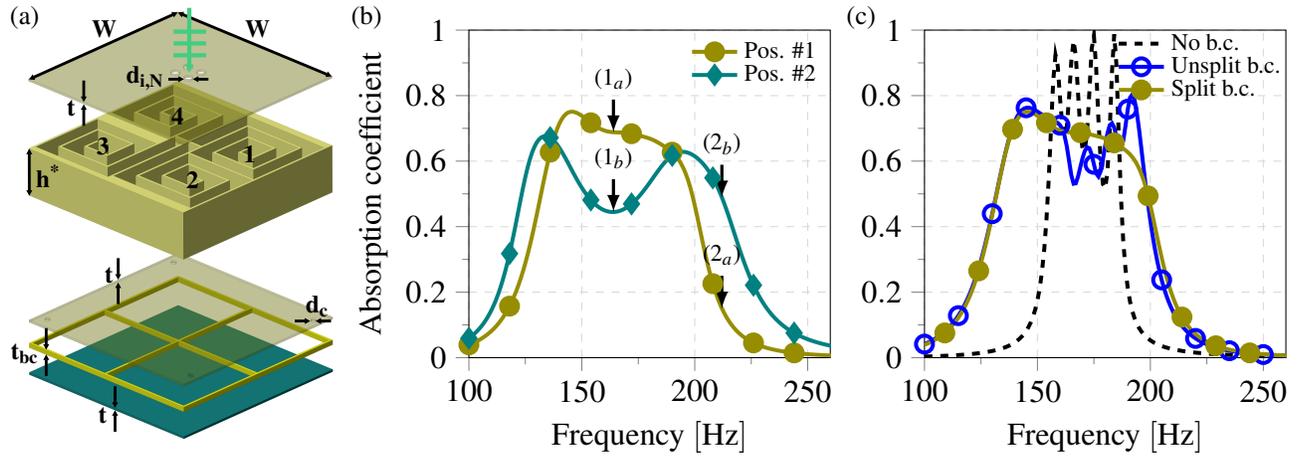}%
\caption{\label{fig:fig3} %
(a) The hybrid metasurface similar to the one in Fig. \ref{fig:fig1a} but with individual back-cavities acquired by splitting the shared back-cavity in equal dimensions. 
(b) The sound absorption dependence on the hole locations shown in Fig. \ref{fig:fig1b}, for the split back-cavity design in (a). $1_{a,b}$ and $2_{a,b}$ indicate two frequencies 164 Hz and 212 Hz, respectively, for both curves to compare the different designs by their pressure field and intensity which are presented in Fig. 4. (c) Absorption spectra for both the designs shown in Fig. \ref{fig:fig1a} and  Fig. \ref{fig:fig3a} (solid lines with markers) along with the case where no back-cavity is included in the design (dashed line). (Symbols for illustration purposes only.)
}%
\end{figure*}%
%
\begin{figure*}[t!]
\begin{subfigure}{0\textwidth}%
\phantomcaption\label{fig:fig4a}%
\end{subfigure}%
\begin{subfigure}{0\textwidth}%
\phantomcaption\label{fig:fig4b}%
\end{subfigure}%
        \includegraphics[scale=1]{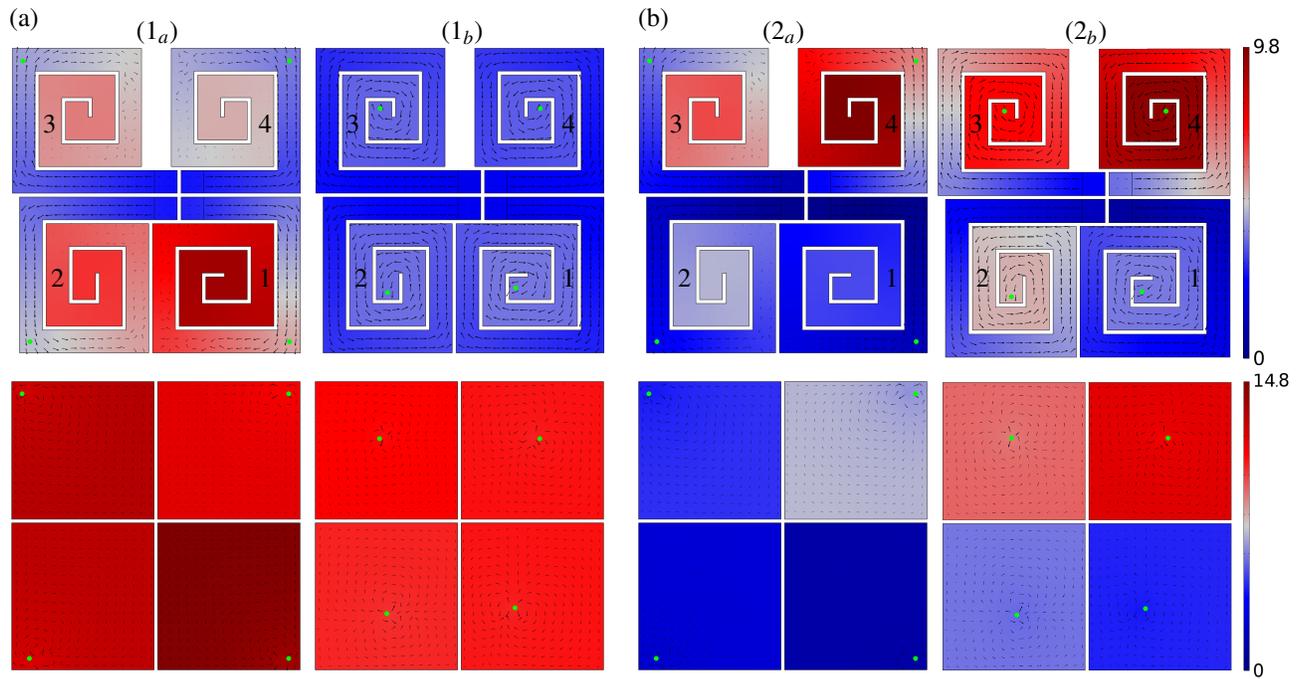}%
\caption{\label{fig:fig4} %
 Acoustic pressure amplitude fields (in Pa) along with the intensity flow fields (indicated by the arrows) at the middle plane through the coiled resonators (upper panel) and back-cavities (bottom panel) of the design depicted in Fig. \ref{fig:fig3a}. The green dots indicate the holes location for the configurations Pos. $\#1$ and Pos. $\#2$ (Fig.1b). The fields are presented at frequencies 164 Hz (a) and 212 Hz (b) indicated by ($1_{a,b}$) and ($2_{a,b}$) respectively in Fig. \ref{fig:fig3b}.%
}%
\end{figure*}%
%
%
%
%
Further improvements of the absorption quality and bandwidth are acquired by amending the back-cavity design through splitting it in four equal parts, one per coil as depicted by Fig. \ref{fig:fig3a}. The new super-cell consists of four paralleled unit-cells, each one consists of a series of two connected resonators. %
The first layer remains the same, the prior individualized four coiled resonators while, the second layer now consisting of four identical resonators %
each of which individually connects to a unit-cell in the first layer without any of the latter sharing any of the former. %
The new design uses the same parametric values from Table \ref{tab:table1} as well as the previously optimized connecting hole locations and dimensions. Considering the fact that the back-cavities are no longer shared among the unit-cells, the connecting hole location previously optimized might no longer be in its optimal position. To check the validity hereof, two of the resulting graphs are depicted in Fig. \ref{fig:fig3b}, from which it can be deduced that the optimal connecting hole location remains to be `Pos \#1'. In fact, while the configuration of Pos.\#2 tend to offer a larger bandwidth, it displays a low absorption coefficient at around 170Hz (below 50\%), while the configuration of Pos.\#1 maintains a relatively high absorption coefficient of 70\% at the same frequency.

A comparison between the shared back-cavity design (Fig. \ref{fig:fig1a}) and the split back cavity design (Fig. \ref{fig:fig3a}) is made for the sound absorption performance. The results are presented in Fig. \ref{fig:fig3c}. The comparison shows that splitting the back cavity has the effect of smoothing the absorption curve while maintaining the desired bandwidth and the overall absorption rate. The smoothing effect improves the overall absorption quality by getting rid of the local minima. %
Besides, to compare our hybrid metasurface capability to that of the classical coiled resonator based metasurfaces, we have added in Fig. \ref{fig:fig3c} the curve of the absorption spectrum for the classical metasurface (w/o back-cavity). %
The addition of the optimized shared back cavity doubles the absorption bandwidth from 155-187 Hz (18.7\%) to 130-200 Hz (42.4\%) at absorption coefficient of 50\%, without passing the minimum local absorption coefficient at 180 Hz in the design w/o back cavity, leading to a doubling of total amount absorbed sound energy even though the maximum absorption is decreased from unity to 80\%. In addition, the smoothing effect caused by the split in the back-cavity creates a further overall increase of the absorption by smoothing away the local minima and so, improving the minimal absorption within the absorption band by roughly 15\% without having any effect on the bandwidth. The split also smooths the highest absorption peak whereby reducing the maximal absorption by 5\% (25\% comparatively to the case without back-cavity). 
%
%
%
%
%
%
%
%
%

Figure \ref{fig:fig3b} contains four arrows labeled 1$_\text{a,b}$ and 2$_\text{a,b}$ at 164 Hz and 212 Hz, respectively, indicating the largest difference in the absorption coefficients for the two different connecting holes locations. These arrows are chosen to display the pressure amplitude distribution within the four coiled cavities and back cavities, displayed in Fig. \ref{fig:fig4}. The objective is to understand the coupling mechanism between the resonators that lead to these absorption coefficient differences. We also indicate in the Fig. 4 the intensity flows (black arrow heads) at the mid-planes of the coils and cavities. At 164 Hz (1$_\text{a,b}$ in Fig. \ref{fig:fig3b}), we observe in Fig. \ref{fig:fig4a} that for the case of connecting holes located at in Pos.\#1 (1$_\text{a}$ ), the four coiled cavities display high acoustic pressure amplitudes along with the back cavities, comparatively to the low amplitude in the coils in case of the situation for Pos.\#2 (fields corresponding to 1$_\text{b}$ ). This lead to a higher absorption coefficient in the spectrum at 164 Hz. In addition, all four coiled cavities are ``active'' in $1_\text{a}$ meaning that nonlocal coupling is acquired, though the largest coiled cavity (coil 1 $\sim$ the bottom right coil) clearly absorbs the largest percentage of the acoustic energy. However, in the case of 1$_\text{b}$ corresponding to the back holes at Pos.\#2, the coiled cavities contribution to the sound absorption capability is almost negligible giving the low pressure amplitude while, the back-cavities display a relatively high pressure amplitude hence being essential to the absorption. For all the observation made from Fig. 4(a), we can conclude that better absorption is attainable when all coiled and back cavities couple and contribute to the overall absorption. 

Furthermore, is the orderly nature of the intensity flow in the coils (1$_\text{a}$ \& 1$_\text{b}$), proof that the design has good impedance matching. %
In the coils of 1$_\text{a}$ the orderly flow becomes disorderly and dies down after the connecting hole, suggesting that the acoustic intensity flow of the acoustic wave after the connecting hole becomes low and eventually negligible likely meaning: almost perfect destructive interference exists in this part of the coil after the hole.

At 212 Hz (2$_\text{a,b}$ in Fig. \ref{fig:fig3b}, the absorption is highest for connecting hole location at Pos. \#2 (2$_\text{b}$) and nonlocal coupling is once more acquired among the paralleled unit-cells. Figure \ref{fig:fig4b} shows high pressure amplitude in the shortest coiled cavities (top ones) which is logical as these operates at higher frequency in comparison to the case of 1$_\text{a,b}$. Furthermore, the top back cavities display a high pressure amplitude for the case of connecting holes in Pos.\#2 (bottom right panel of Fig.4(b)) in comparison to Pos.\#1 (bottom left panel of Fig.4(b)). The combination of resonating back cavities and coiled cavities (highlighted by the high pressure amplitude) lead to an enhancement of the absorption coefficients. 
Proper impedance matching still exists as shown by the intensity flow in Fig. \ref{fig:fig4b} as per the same reasoning previously mentioned. %

So, the aforementioned comparisons make two things clear: first, the back-cavities are the stronger absorbers, and second, the highest rate of absorption is acquired when both the coils and the back-cavities work in tandem. %

In summary, a new sub-wavelength (50 mm$\sim\nicefrac{\lambda}{51}$) thick design has been introduced, consisting of four coiled resonators in parallel and each independent resonator being in series with another resonator, bringing about low-frequency broadband absorption from 130-200 Hz at 50\% absorption. %
The design doubles the bandwidth of the classical super-cell design of only four paralleled HR, even though it's peak absorption is lowered by 25\%. %
The improvements are attributed to the combination of the proper impedance matching and the nonlocal coupling seen from the intensity flow and pressure fields respectively, and the back-cavities are responsible for the majority of the absorption within the design. %
The latter being easily introduced to other designs, makes it a good candidate to improve future designs. %
%
%
%
%
%
%
\begin{acknowledgments}%
The authors acknowledge the joint research laboratory MOLIERE and la R\'{e}gion Grand Est for supporting this work.
\end{acknowledgments}%

\section*{Conflict of interest}%
The authors have no conflicts of interest to disclose.%

\section*{Data Availability Statement}%
Restrictions apply to the availability of these data, which were used under license for this study. Data are available from the authors upon reasonable request.%
\nocite{*}%
\section*{References}%
\providecommand{\noopsort}[1]{}\providecommand{\singleletter}[1]{#1}%
\end{document}